\documentclass[12pt]{iopart}
\usepackage{iopams}
\usepackage{setstack}

\usepackage[usenames]{color}

\usepackage{graphicx}
\usepackage[noadjust]{cite}
\usepackage{hyperref}

\begin{document}
\comment 
[Comment on `Oxygen vacancy-induced magnetic moment in\dots Li$_{2}$CuO$_{2}$']
{\vspace{-1mm}
Comment on `Oxygen vacancy-induced magnetic moment in edge-sharing
CuO$_{2}$ chains of Li$_{2}$CuO$_{2}$'}
\vspace{-2mm}
\author{R~O~Kuzian$^{1,2}$, R~Klingeler$^{3}$, 
W~E~A~Lorenz$^2$, N~Wizent$^{2,3}$,
S~Nishimoto$^{2,4}$, U~Nitzsche$^2$, H~Rosner$^5$, D~Milosavljevic$^5$, L~Hozoi$^2$, R~Yadav$^2$, 
J~Richter$^6$, A~Hauser$^6$,   
J~Geck$^{4}$, R~Hayn$^7$, V~Yushankhai$^{8}$, L~Siurakshina$^{8}$, C~Monney$^{9}$, T~Schmitt$^{10}$, 
J~Thar$^{11}$,
G~Roth$^{11}$, T~Ito$^{12}$, 
H~Yamaguchi$^{12}$, M~Matsuda$^{13}$, S~Johnston$^{14}$, 
J~M\'alek$^{15}$,
and  S-L~Drechsler$^{2}$
}
\address{$^{1}$ Inst.\ f.\ Problems of Mat.\ Science NASU, Krzhizhanovskogo
3, 03180 Kiev, Ukraine}
\address{$^{2}$ Inst.\ f.\ Solid State Res.\ \&
Mat.\ Science, IFW-Dresden, D-01169 Dresden, 
Germany}
\address{$^{3}$ Kirchhoff Institute f.\ Physics,  Heidelberg University,
D-69120 Heidelberg, Germany}
\address{$^{4}$ 
Department of Physics, TU Dresden, Dresden, Germany}

\address{$^{5}$ Max-Planck-Institute for Chemical Physics of Solids, Dresden, Germany}
\address{$^{6}$ Institute for Theoretical Physics, University of Magdeburg, Magdeburg, Germany}
\address{$^{7}$ Aix-Marseille Universit\'e, IM2NP-CNRS UMR 7334, 
13397 Marseille, France}

\address{$^{8}$ Joint Institut for Nuclear Physics, Dubna, Moscow region, 
Russia}
\address{$^{9}$ Department of Physics, University of Zurich, 8057 Zurich, Switzerland}
\address{$^{10}$ Paul Scherrer Institute PSI CH-5232 Villingen,  Switzerland}
\address{$^{11}$ RWTH Aachen University, Institute for Crystalography, Aachen, Germany}
\address{$^{12}$ Nat.\ Inst.\  Adv.\ Indust.\ Sci.\ \& Techn.\ (AIST), Tsukuba, Ibaraki 305-8562, Japan}
\address{$^{13}$ Neutron Scattering Div., Oak Ridge National Lab., Oak Ridge, TN 37831, USA}
\address{$^{14}$ Dept.\ of Phys.\ \& Astron., The University of Tennesee, Knoxville, TN 37996, USA}
\address{$^{15}$ Inst.\ of Physics, ASCR, Na Slovance 2, CZ-18221 Prague, Czech Republic}
\ead{s.l.drechsler@ifw-dresden.de}
\begin{abstract}
In a recent work devoted to the magnetism of  Li$_{2}$CuO$_{2}$, Shu \emph{et
al.} [New J.\ Phys.\ 19 (2017) 023026] have proposed a ``simplified"
unfrustrated microscopic model that differs considerably from 
the models 
refined through decades of prior work.  We show that the 
proposed model is at odds
with known experimental data,
including the reported magnetic susceptibility $\chi(T)$ data up to 550~K.
Using an 8$^{\rm th}$ order high-temperature
expansion 
 for $\chi(T)$, we show that the
experimental 
data for Li$_{2}$CuO$_{2}$ are consistent with the prior model
derived from inelastic neutron scattering (INS)  studies. We
also establish the $T$-range of validity for a 
Curie-Weiss law for the 
real frustrated magnetic system.  We argue that the knowledge of the long-range ordered
magnetic structure for $T<T_N$ and of $\chi(T)$ in a restricted $T$-range 
provides insufficient information to extract all 
of 
the relevant couplings in
frustrated magnets; the saturation field and INS data must also be used to
determine several exchange couplings, including the weak
but decisive
frustrating antiferromagnetic (AFM) interchain couplings.

\end{abstract}
\vspace{-0.5cm}


\submitto{\NJP}

\maketitle
Li$_2$CuO$_2$ takes a special place among the still increasing family of
frustrated chain compounds with edge-sharing CuO$_4$ plaquettes and a
ferromagnetic (FM) nearest neighbor (NN) in-chain coupling $J_1$
\cite{Drechsler07}.  This unique position is due to its ideal planar CuO$_2$
chain structure and its well-defined ordering characterized by a 3D
Ne\'el-type arrangement of adjacent chains whose magnetic moments are aligned
ferromagnetically along the chains ($b$-axis). Li$_2$CuO$_2$ is
well studied in both experiment and theory
(see e.g.\ 
\cite{Sapina90,Graaf02,Yushankhay02,Malek08,Lorenz09,Money13,Nishimoto15,Johnston16,Money16,Mueller17}) 
and serves nowadays as a
reference system for more complex and structurally less
ideal systems. In particular, it is accepted in the quantum magnetism community
that the leading FM  coupling is the NN inchain coupling
$J_1$. ($J_1$ is also dominant but antiferromagnetic (AFM) in the special 
spin-Peierls 
case of CuGeO$_3$ \cite{Geertsma96}.)
There is
always also a {\it finite frustrating} AFM
next-nearest neighbor (NNN) coupling $J_2>0$,  see figure 1, left.
This inchain frustration is
\begin{figure}[b]
\centering
\includegraphics[width=0.34\columnwidth]{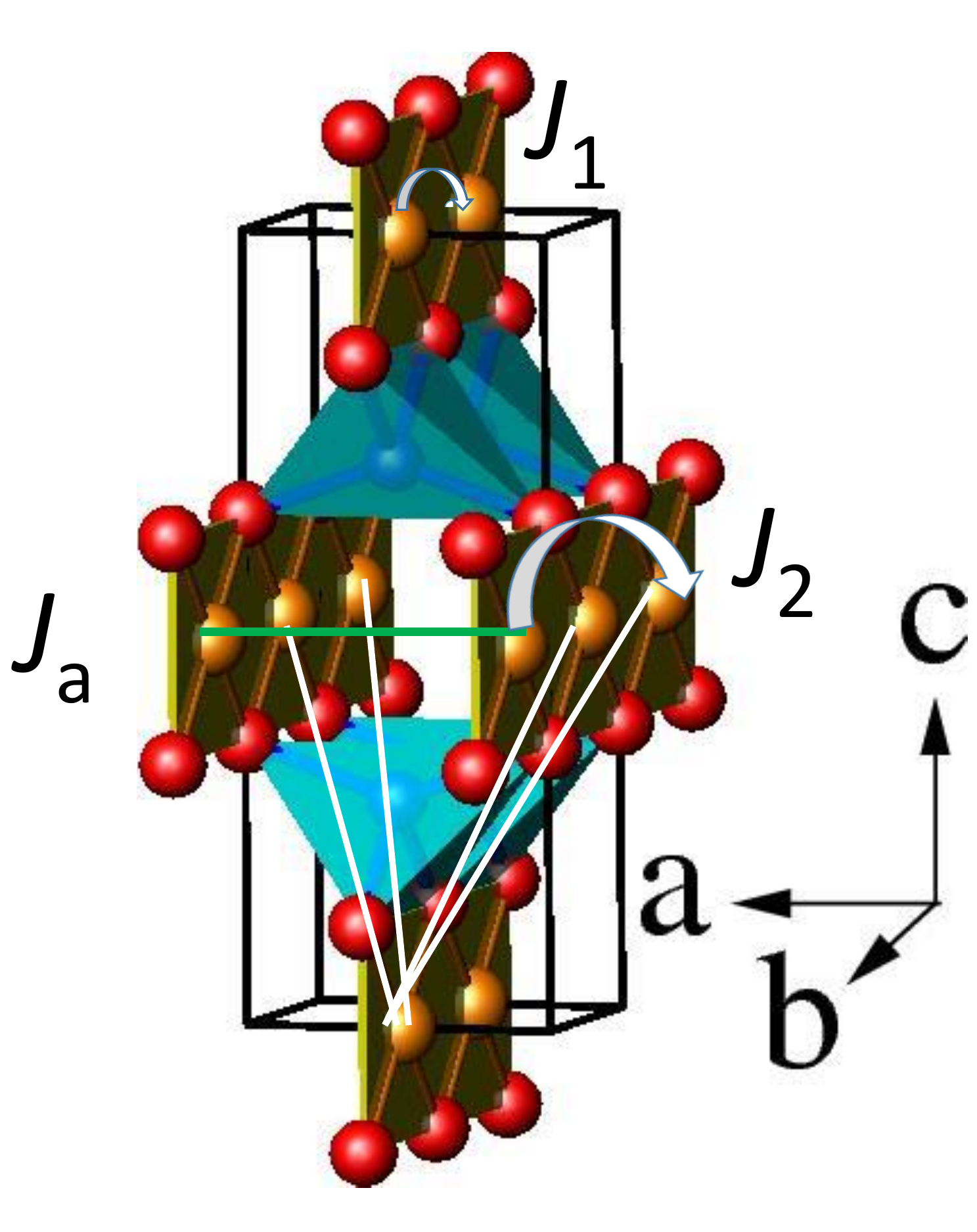}
\includegraphics[width=0.65\columnwidth]{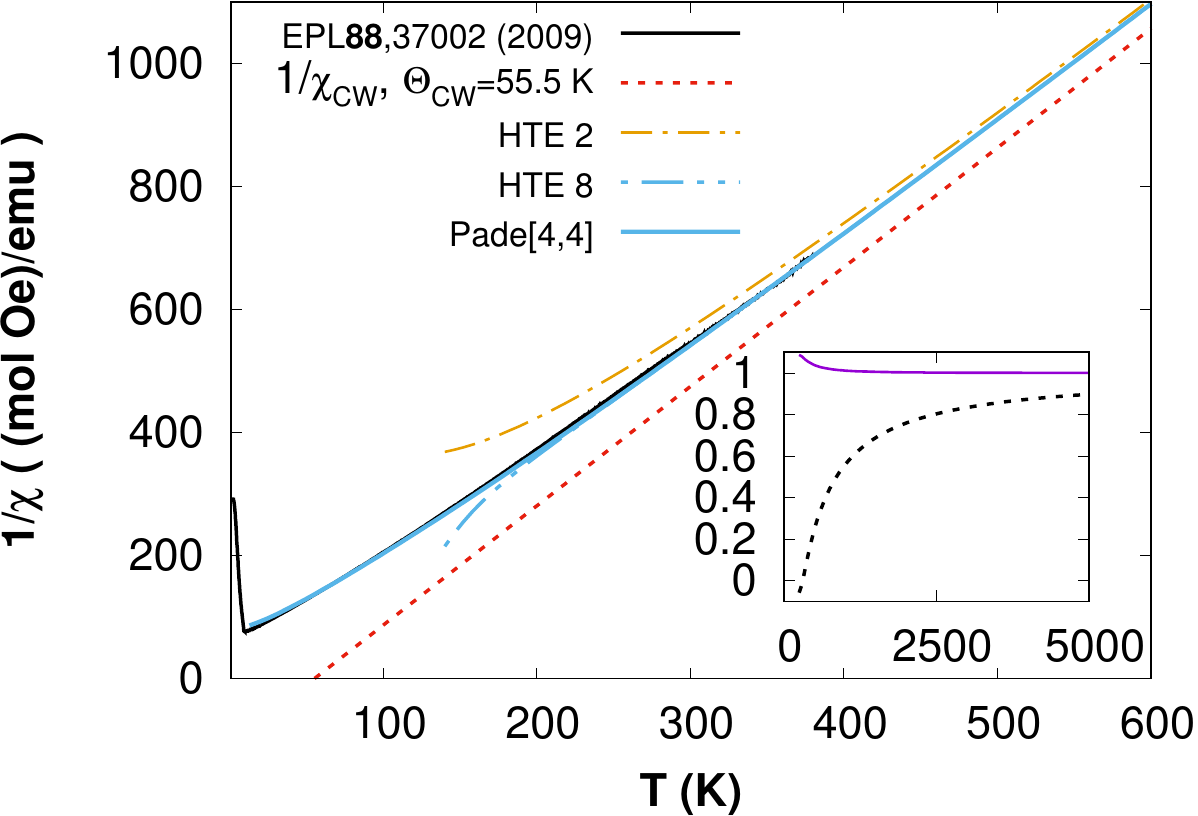}
\caption{Left: the crystal structure of real Li$_2$CuO$_2$ and the 
 main exchange couplings under debate (White lines:  1/4 of the skew
 AFM frustrating 
 IC's $J_5$ and $J_6$ between adjacent 
 chains responsible for the FM inchain ordering
 but ignored in \cite{Shu17}, similarly as the generic AFM NNN inchain coupling 
 $J_2$. There, the FM NN  intrachain coupling $J_1$ is underestimated by a factor of four
 (see table 1).
  Green line: the 
 weak NNN IC $J_3\equiv J_a$  claimed to be dominant and FM  by Shu {\it et al.}\cite{Shu17}). 
Right: \textbf{Main:} The inverse spin susceptibility
for a magnetic field along the $a$ axis of Li$_{2}$CuO$_{2}$
(the sample that was used for INS studies, cf.\ figure 4
of \cite{Lorenz09})  fitted by the eight-order high-$T$
expansion expression (\ref{eq:fitchi}). 
\textbf{Inset:} Convergence of the two pseudo-CW parameters 
to their high-$T$ asymptotic values: $C^*(T)/C$ (solid line),
$\Theta^*_{\rm \tiny CW}(T)/\Theta_{\rm \tiny CW}$ (dashed line). 
}
\label{fig:fig1} 
\end{figure}
quantified by $\alpha=J_2/\left|J_1 \right| $.  
In the present case, and in that of the related Ca$_2$Y$_2$Cu$_5$O$_{10}$, 
there are  only frustrating AFM {\it interchain} couplings (IC) 
with adjacent 
chains {\it shifted} by half a lattice constant $b$.
In this lattice structure there 
is {\it no} room for unfrustrated perpendicular IC.
This AFM IC with NN {\it and} NNN components
plays a decisive role in the stabilization of the FM alignment of the magnetic 
moments along the chain direction.  
Although weak at first glance, with eight NN and NNN it is nevertheless
 significant enough (by a factor of two) to 
prevent a competing non-collinear spiral type ordering 
in Li$_2$CuO$_2$ (with 
the frustration ratio $\alpha > 1/4$),  
as often observed for other members of this family with unshifted chains
\cite{Drechsler07}.  All these well-established features were
practically excluded by Shu, Tian, Lin {\it et al.} (STL) \cite{Shu17}, 
proposing instead (i) 
a very nonstandard {\it unfrustrated} model (dubbed hereafter
as  STL-model) with comparable couplings in all 
directions, and where the leading FM coupling is given by an unphysically 
large NNN FM IC
$J_a$ (denoted as $J_3$ therein) perpendicular to the chains in the basal 
$ab$-plane ($J_a$=$-103$~K for stoichiometric and $-90$~K in
the presence of O vacations in Li$_2$CuO$_{2-\delta}$ with $\delta=0.16$). 
(ii) The coupling between the NN chains, as derived from the inelastic
neutron scattering (INS) 
data \cite{Lorenz09} and in 
qualitative accord with the results of LDA+$U$ calculations \cite{Malek08}, 
has been ignored and 
replaced {\it ad hoc} by an artificially large, ``effective"  {\it non-frustrated } 
AFM IC
$J^{\prime\prime}$ (see the right of figure 9 in \cite{Shu17} with a 4-fold coordination) {\it absent} in real Li$_2$CuO$_2$.

In the present Comment, we 
show that this parametrization is a direct 
consequence of an incorrect analysis of their susceptibility $\chi(T)$ data
in addition to ignoring the highly dispersive magnon mode and its local softening
observed by INS. We admit that the very question about an influence of O vacancies on the
magnetic properties of Li$_2$CuO$_2$ raised by Shu \emph{et al}. \cite{Shu17} is
interesting and should be studied; however, this must be done
within a proper analysis based on a {\it realistic} phenomenological model 
reflecting the established large $|J_1|$  values exceeding 200~K
\cite{Lorenz09} and {\it excludes} a simple Curie-Weiss (CW) law below 1000~K.
In  this context we
mention similar mistakes made in the literature,
where even an artificial AFM $\Theta _{CW}<0$ has been found 
\cite{Sapina90,Ebisu98,Sredhar88,Boehm98}
prior to 2009, when the large value of $J_1$ was not yet known.

Li$_{2}$CuO$_{2}$ is a 
frustrated quasi-1D system that has been well
studied during the last decades. 
The first two rows of table 1.
(see table 6 of\cite{Shu17}) 
provide the $J$'s suggested
by Shu \emph{et al.} from their qualitative
simulation of the magnetic  ordering
and an analysis of the measured $\chi(T)$ for  two 
samples with different O 
content.
The main striking difference between these sets from all previous 
ones is the {\it absence} of both magnetic frustration and of 
the quasi-1D regime with a dominant $J_1$ realized in 
all edge-sharing  CuO$_2$ chain compounds. 
Moreover, the proposed sets are evidently at odds
with the results of two
INS studies \cite{Lorenz09,Boehm98}.
The set derived from the INS in the $bc$ plane \cite{Lorenz09} 
see the last row in table 1
(and table 6 of \cite{Shu17}) 
does
explain the 
$\chi(T)$ data for $T>T_{N}$, 
especially 
when supplemented by a weak AFM IC  $\parallel$ to
the $a$ axis (in accord with the reported weakly dispersing magnon 
in that direction   
\cite{Boehm98,Lorenz11}).

We consider the Heisenberg spin-Hamiltonian 
\begin{equation}
\hat{H}=\frac{1}{2}\sum_{\mathbf{m,r}}J_{\mathbf{r}}
 \hat{\mathbf{S}}_{\mathbf{m}}\cdot\hat{\mathbf{S}}_{\mathbf{m+r}}
 -g_{a}\mu_{B}H_{z}\sum_{\mathbf{m}}\hat{S}_{\mathbf{m}}^{z},
\label{eq:H}
\end{equation}
where $\mathbf{m}$ enumerates the sites in the magnetic (Cu) lattice
(see figures \ref{fig:fig1} and 9 in  \cite{Shu17}), $J_{\mathbf{r}}$ is  the interaction 
of a pair of spins $\hat{\mathbf{S}}_{\mathbf{m}}$ and 
$\hat{\mathbf{S}}_{\mathbf{m+r}}$. Note the different 
notation  
\ \cite{Takeda95,Carter96,Boehm98}, where the same interaction is denoted
$-2J_{\mathbf{r}}$. 
The form of 
equation (\ref{eq:H}) 
implies positive (negative) signs for AFM (FM) couplings. 
Shu \emph{et al.} \cite{Shu17} use the same notation in tables 5 and 6,
but use the wrong signs in equations (8) and (10).
A small anisotropy of the couplings seems to be unimportant for the 
analysis of $\chi(T)$ and is ignored here.
The magnetic field $H_{z}$
is directed along the easy axis (i.e.\
the crystallographic $a$-axis in Li$_{2}$CuO$_{2}$). 
Finally, $\mu_{B}$ is the Bohr magneton and $g_{a}$
is the gyromagnetic ratio for this direction. 
\begin{table}
\noindent
\hspace{-2cm}
\caption{\label{tab:Ji} Exchange sets proposed for the ``effective" unfrustrated 
STL-model
and our frustrated one for real Li$_{2}$CuO$_{2}$ in the notations of \cite{Shu17}, 
see figures 9(b) and (a), respectively,  therein. $J_a$=$J[1,0,0], J_5$=$J[\case{1}{2},\case{1}{2},\case{1}{2}]$, 
and $J_6$=$J[\case{1}{2},\case{3}{2},\case{1}{2}]$
in crystalographic notations.}
\begin{indented}
\item[]
\begin{tabular}{@{}lllllllll}
\br
$J_{i}/k_{B}$ (K) & $J_{1}$ & $J_{2}$ & $J_{3}\equiv J_a$ &
 $J_{4}$ & $J_{5}$ & $J^{\prime\prime}$ & $J_{6}$ & $\Theta _{\rm \tiny CW}(K)$\\
\mr
$z_{i}$ & 2 & 2 & 2 & 4 & 8 & 4 & 8 & \\
\mr
\cite{Shu17}($\delta\sim 0.16$)& -61 & 0 & -90 & 0 & - & 62 & 0 &  14\\
\cite{Shu17} ($\delta\sim 0$)& -65 & 0 & -103 & 0 & - & 71.2  & 0 & 13\\
\mr
\cite{Lorenz09} & -228 & 76 & - & - & 0 & - & 9.04 & $\stackrel{>}{\sim}$50\\
\cite{Lorenz11},present & -230 & 75 & 4.8 & 1.6 & 0 & - & 9 & 56\\
\br
\end{tabular}
\end{indented}
\end{table}
 
Shu {\it et al.} 
use two approaches when
analyzing 
$\chi(T)$
of Li$_{2}$CuO$_{2}$. First, they fit it
in the range 250~K$<T<550$~K using a CW-law
\begin{equation}
\chi_{CW}(T)=\chi_{0}+\frac{C}{T-\Theta_{CW}}\label{eq:chiCW}.
\end{equation}
They then extract three effective couplings  instead of six original ones
by fitting $\chi (T)$ with 
an 
RPA-like expression derived for quasi-1D systems. Note that Shu \emph{et al.} give 
an obviously erroneous form 
in their equation~(7) with
the factor 
$\left[1-2\left(z^{\prime}J^{\prime}+z^{\prime\prime}J^{\prime\prime}\right)\right]$,
being a 
sum of the dimensionless value 1 and a
value with the dimension of energy. No 
figure is shown for
$\chi(T)$
fitted
by their curve, so it is impossible to evaluate
their fit, nor to estimate its quality and 
validity range. However, in \cite{Takeda95} (reference  33 of their paper), 
the correct expression reads:
\begin{equation}
\chi_{\mathrm{q1D}}  =  \frac{\chi_{\mathrm{1D}}}{1+
 \displaystyle\frac{\left(z^{\prime}J^{\prime}+z^{\prime\prime}
 J^{\prime\prime}\right)\chi_{\mathrm{1D}}}{Ng^{2}\mu_{B}^{2}}},
\quad
\chi_{\mathrm{1D}}  =  \frac{Ng^{2}\mu_{B}^{2}}{4k_{B}T}\left(1-
\frac{J}{2k_{B}T}\right),
 \label{eq:quasi1D}
\end{equation}
where $J$=$J_{1}$ is the inchain coupling and $J^{\prime}$=$J_{3}$, $J^{\prime\prime}\sim 2J_{5}$
are IC's  with $z^{\prime}$=$2$, $z^{\prime\prime}$=$4$ the corresponding numbers 
of neighbors (see figures 9(b) in \cite{Shu17}
for a simplified structure, which differs from the real one, see figure.~9(a) therein). 
In equation 
(\ref{eq:quasi1D}) 
we have accounted for the different notations of the exchange couplings
between \cite{Shu17} (the same as ours) and \cite{Takeda95}.
Note that the quasi-1D regime assumed in equation (\ref{eq:quasi1D})
implies 
$J\gg\left(z^{\prime}J^{\prime}+z^{\prime\prime}J^{\prime\prime}\right)$,
which is obviously {\it violated} by the STL model. 
We recall that a
CW-law  \emph{exactly} reproduces the high-$T$ 
behavior of the spin susceptibility of any system 
described by the Heisenberg Hamiltonian (\ref{eq:H}) with the CW temperature
$\Theta_{\rm \tiny CW}$ 
\begin{equation}
\Theta _{\rm \tiny CW}=-\frac{S\left(S+1\right)}{3k_{B}}\sum_{i}z_{i}J_{i}=-\frac{1}{4k_{B}}\sum_{i}z_{i}J_{i} \ .
\label{eq:TCW}
\end{equation}
Equation (\ref{eq:TCW})  is the \emph{exact} 
result of a high-$T$ expansion (HTE) of the susceptibility
\cite{Opechowski37} (see equation (27) of section IV.B in \cite{Schmidt01},
see also \cite{Lohmann14},
and references
therein), which is valid for \emph{any} Heisenberg system. 
As 
mentioned above, the expression for $\Theta _{\rm \tiny CW}$ given in
equation (10) of \cite{Shu17} has a wrong sign. It gives
positive (negative) contribution for AFM (FM) interactions
in conflict with the physical meaning of  $\Theta _{\rm \tiny CW}$.  

Now we show that $\chi_{\mathrm{q1D}}$ also obeys
a CW-law for large enough $T$. We recast equation (\ref{eq:quasi1D}) in the form
\begin{equation}
\fl \frac{C}{\chi_{\mathrm{q1D}}T}=\frac{1}{1-\displaystyle\frac{zJ}{4k_{B}T}}
 +\frac{z^{\prime}J^{\prime}+z^{\prime\prime}J^{\prime\prime}}{4k_{B}T} 
  =  1+\frac{zJ+z^{\prime}J^{\prime}+z^{\prime\prime}J^{\prime\prime}}{4k_{B}T}
  +\sum_{n=2}^{\infty}\left(\frac{zJ}{4k_{B}T}\right)^{n},
  \label{eq:recast} 
\end{equation}
where $C$ denotes
the Curie constant, $z=2$ is the in-chain coordination number, and
\begin{equation} 
\chi_{\mathrm{q1D}}  =  \frac{C}{T-\Theta_{CW}}\left[1+
\mathcal{O}\left(\frac{zJ}{4k_{B}T}\right)\right],
\quad C=\frac{Ng^{2}\mu_{B}^{2}}{4k_{B}} \  .
\label{eq:CWq1D}
\end{equation}

The $\Theta_{\rm \tiny CW} =+14$~K for the 
STL-model
given by equation (\ref{eq:TCW}) should coincide with the value
obtained by the CW fit presented in tables  3 and 4 of \cite{Shu17}, 
provided the CW fit is justified for the chosen range of $T$.
   We stress that a 
   value of $\Theta _{CW} >0$ \emph{doesn't} cause a 
   divergence of $\chi (T)$ at $T=\Theta _{CW}$, in contrast to the
   erroneous claim of Shu \emph{et al.} in 
   speculations after their 
   equation (10).  
   In general, a divergence of
   the susceptibility $\chi ({\bf Q}_0, T\to T_0)$ means the emergence of a long-range
   order in a magnetic system characterized by a spontaneous magnetization
   $m({\bf Q}_0)$ for $T<T_0$. A ferrimagnetic (FIM) 
   and a FM
   ordering correspond
   to ${\bf Q}_0$ at the center of 
   Brillouin zone (BZ) and to 
   the uniform component of $\chi (0,T)\equiv \chi(T)$, respectively.  
   Note that a FIM 
   ordering
   and the divergence of $\chi (T\to T_C)$ may occur for
   systems with purely AFM couplings and \emph{negative}  $\Theta _{CW}$
   due to the geometry of the spin arrangement  
   (see e.g.\ figure 4 of  \cite{Kuzian16}).
   A ${\bf Q}_0 \neq (0,0,0)$ corresponds to a helimagnetic or to
   an AFM ordering. Then, the uniform 
   $\chi (T)$ remains 
   finite at $T_0$. An
   AFM ordering corresponds  
   to  ${\bf Q}_0$ located at the edge of the BZ. 
  This is the case
   for the prior model of Li$_2$CuO$_2$, \emph{as well as} for the 
   STL-model. 
   The range of validity of the 
   CW-law for $\chi_{\mathrm{q1D}}$ [equation~(\ref{eq:quasi1D})] is 
\begin{equation}
k_{{\rm \small B}}T\gg zJ/4 .
\label{eq:range1D}
\end{equation}
As 
already noted, the approximate expression equation (\ref{eq:quasi1D})
is relevant for a quasi-1D system, where the inchain
$J$'s  dominates. This is true also for the condition (\ref{eq:range1D}).
Let us establish now a general condition for the applicability of a 
CW-law. For this aim
it is convenient to use the
inverted \emph{exact} HTE \cite{Johnston00} for spin-1/2 systems
with equivalent sites
\begin{equation}
\fl \frac{C}{\chi T}=1+\frac{D_{1}}{T}+\frac{D_{2}}{T^{2}}+\cdots,\quad 
D_{1}=\frac{1}{4}\sum_{i}\frac{z_iJ_{i}}{k_{B}},\quad 
D_{2}=\frac{1}{8}\sum_{i}z_i\left(\frac{J_{i}}{k_{B}}\right)^{2},
\label{eq:JohnHTE}
\end{equation}
(cf.\ equations 5(a) and (b) in \cite{Johnston00}). Thus, a 
CW-law with $\Theta _{CW}=-D_1$ is valid in the range
\begin{equation}
T\gg\sqrt{D_{2}}> 
\textcolor{black}{\max\left(\frac{z_i\left|J_{i}\right|}{4k_{{\rm \tiny B}}},
\left|\Theta _{CW}\right|\right).}\label{eq:rangeCW} 
\end{equation}
From this expression it is clear that for systems where both FM and
AFM couplings are present, the CW behavior is valid at
$ T\gg \left|\Theta_{\rm \tiny CW}\right|$. For the unfrustrated STL-model 
(the first row in table 1), the condition (\ref{eq:rangeCW})
reads $T\gg70$~K, while for the prior INS based model (the last row in 
table 1) it is
$T\gg120$~K. One should also take into account that the convergence of the HTE 
is slow. To show that the  exchange values determined from the INS are 
compatible with the $\chi (T)$ data, we reproduce in figure \ref{fig:fig1} 
the data measured on the same sample
(i.e.\ with the same O content) used for the
INS studies (see figure 4 of \cite{Lorenz09}). 
We have fitted the data in the range $340<T<380$~K with the expression 
\begin{equation}
\chi (T)=\frac{N_Ag_a^{2}\mu_{B}^{2}}{k_{B}}\chi _8(T), 
\label{eq:fitchi}
\end{equation}
where $\chi _8(T)$ is the eighth-order HTE obtained by the method and 
programs published in \cite{Schmidt11}\footnote{The corresponding HTE package
used in [25] we have employed obtaining the results shown here,too. It is
available at
 http://www.uni-magdeburg.de/jschulen/HTE/ }, 
and $N_A$ is  Avogadro's number. 
The HTE program of \cite{Schmidt11} we have used here
can only treat systems with four different 
exchange couplings,only, so we decided
to adopt one effective coupling in the $a$ direction 
$J_a=J_3+2J_4$. Only \emph{one} 
parameter $g_a$ was adjusted during the
 fit despite the small background susceptibility $\chi_0$ which was set to zero
 for the sake of simplicity and its insensitivity to our fit. The [4,4]-Pad\'e approximant for 
the HTE fits the data well down to $T\sim 20$~K 
with the reasonable value $g_a=2.34$.
The inset shows $C^*(T)$ and $\Theta^*_{\rm \tiny CW}(T)$, 
the two parameters of the pseudo-CW-law 
(cf.\ figure 4 of \cite{Lorenz09}), which is given by a tangent to 
the $1/\chi(T)$ curve at a given $T$.
The pseudo-Curie ``constant" 
$C^*(T)$ exceeds its asymptotic value for all $T$. Hence,
it can not be
used 
to 
extract
the number of spins in the system. The $C$ from a more elaborate HTE is 
much better suited for this aim. 
For example, a clear crossover between a CW ({\it without} any divergence at 
$\Theta_{\rm \tiny CW}$=+39~K) and a pseudo-CW regimes 
has been observed 
recently near 150~K
for CuAs$_2$O$_4$, see figures 8 and 9 in \cite{Caslin14}.

We note that the difficulties encountered at intermediate $T$ when
avoiding the use of a HTE as demonstrated above
can be circumvented by
applying a strong magnetic field (up to saturation at $H_s$ where the 
AFM IC is suppressed). 
At very low $T$ and in the isotropic limit, one has a very simple 
but useful relation
\begin{equation}
  g\mu_{\mbox{\tiny B}}H_s+2J_a+4k_B\Theta_{\rm \tiny CW}=2\mid J_1 \mid (1-\alpha) \ .
  \label{satfield2}
  \end{equation}
 The quantities on the r.h.s.\ can be deduced from the INS
 data \cite{Lorenz09,Lorenz11}. Indeed, with $J_1= -230$~K, $\alpha=0.326$,
 $H_s$=55.4~T for $H||b$, $g_b=2.047$ \cite{Nishimoto11},  
 and $J_a\approx 8$~K, 
 we arrive nearly at the same $\Theta_{\rm \tiny CW}=54$~K as in the HTE
 analysis of $\chi(T)$. 
For collinear systems, $H_s$ 
 is related to the inter-sublattice couplings
(see equation (1) in \cite{Nishimoto11}, valid at $T$=0)
  \begin{equation}
  g\mu_{\mbox{\tiny B}}H_s=N_{IC}(J_{IC}+J_{IC}^{\prime}) \ ,
  \label{satfield}
  \end{equation}
  where $N_{\rm \small IC}$=8 is the number of nearest IC neighbors,
  $J_{\rm \small IC}$=$J_6$, and $J_{\rm \small IC}^{\prime}$=$J_5$.
  Equation (12) gives $(J_{\rm \small IC}+J_{\rm \small IC}^{\prime})/k_B$=9.5~K,
  close to the value from
  our model. Conversely, equation (\ref{satfield}) 
  gives $H_s=4J^{\prime\prime}/(g_a\mu _B)\sim 145$~T
for the STL model, 
even for the unrealistic
value $g_a=2.546$ (table 3 in \cite{Shu17}), which much exceeds
the observed 55~T. Thus, Shu {\it et al}.'s $J^{\prime\prime}$ value is
at odds with the high-field data reported in 2011 \cite{Nishimoto11}.
Shu \emph{et al.} have also ignored the $T$-dependence of the RIXS
(resonant inelastic x-ray spectroscopy) 
spectra of Li$_2$CuO$_2$ reported in 2013 \cite{Money13}, which
were explained within a $pd$-model consistent with the INS data. The 
detection of an intrachain Zhang-Rice singlet exciton at $T>T_N$
and significant increase of its weight at 300~K evidence
the increase of AFM correlations in the chain 
and cannot be explained without
a frustrating AFM intrachain coupling $J_2$. 

Next, we remind the reader of some microscopic 
insights into the magnitude of the
exchange couplings for Li$_2$CuO$_2$. Empirically, a dominant FM $J_1$ \cite{Lorenz09,Lorenz11}
and the minor role of $J_a$ \cite{Lorenz09,Lorenz11,Boehm98} follow directly from  the 
observed strong (weak) dispersion of the magnons along the $b (a)$-axis, respectively. 
 Microscopically, 
the large value of $J_1$ as compared with -100~K suggested in 
\cite{Mizuno98} is a consequence of an enlarged direct FM exchange $K_{pd}$  
between two  holes 
on Cu and O sites within a CuO$_4$ plaquette 
and the non-negligible Hunds' coupling between two  holes in different O 2$p_x$
and 2$p_y$ orbitals in edge-sharing CuO$_2$ chain compounds. 
The former is well-known from quantum chemistry studies (QCS)
of corner-sharing 2D cuprates having similar CuO$_4$ plaquettes with 
$K_{pd,\pi}\approx -180$~meV for $\Phi=180^\circ$ Cu-O-Cu bond angles 
\cite{Hybertsen92}. Now we   
confirm similar values of the two $K_{pd}$ interactons
in Li$_2$CuO$_2$ using QCS \cite{Yadav17}, 
which yielded for $\Phi \approx 90^o$:
$K_{p_xd}\approx K_{p_xd,\pi/2}=K_{p_yd,\pi/2}\approx K_{p_y,d}\sim -100 \ 
\mbox{meV within}\ K_{p_xd}+K_{p_yd} 
< K_{pd,\pi} < -\sqrt{K^2_{p_xd,\pi/2}+K^2_{p_yd,\pi/2}} $, 
(ignoring a weak crystal field splitting
of the two O onsite energies). Thus, our two $K_{p_xd}\neq K_{p_yd}$ values being naturally slightly different
are more than twice as large as -50~meV for both direct FM couplings 
{\it ad hoc}
adopted in \cite{Mizuno98,Geertsma96} 
in the absence of QCS calculations  
for Li$_2$CuO$_2$ and CuGeO$_3$ \cite{Braden96,Geertsma96} 20 years ago.
Concerning $J_a$, QCS yield a small direct FM $dd$
contribution $\approx$ -1~K, while our DFT-derived  Wannier-functions point 
to a tiny AFM value
of about 5~K. A more direct 
calculation within the LDA+$U$ 
scheme
for large supercells is in 
preparation. Anyhow, even with possible error bars of the DFT and QCS
calculations, a value of $J_a$
 exceeding $\pm 10~$K is not expected. Hence, a much
 larger $J_a \approx -103$~K as suggested by Shu {\it et al.} is either an 
 artifact from the constructed
 ``simplified" non-frustrated STL model and/or a consequence of the improperly 
 analyzed $\chi(T)$-data, as explained above. 
 
Thus, we have shown that the ``effective"
 unfrustrated STL-model proposed by Shu {\it et al.} together with its 
 doubtful analysis of $\chi(T)$ are at 
 odds with our current understanding of real
 Li$_2$CuO$_2$ and will only confuse readers. It may, however, be of 
 some academic interest to illustrate the crucial effect of intra- 
 and IC frustrations evidenced by a recent INS study.
 In case of finite NNN exchange $J_2$ with $\alpha > \alpha_c$=1/4, as evidenced 
 by the minimum of the magnon dispersion near the 1D propagation vector, 
 there would be an inphase or antiphase ordering of non-collinear spirals 
 for FM (AFM) perpendicular IC $J^{\prime\prime} $\cite{Zinke09}. Such a
 situation would also be realized  for an STL-model corrected by $\alpha > \alpha_c$, 
 in sharp contrast to the observed
 collinear magnetic moments aligned ferromagneticaly along the  chain direction 
 with weak quantum effects.
 In this context, we note that we know of no example of a frustrated model 
 that can be replaced comprehensively
 by a non-frustrated effective model 
 as Shu {\emph{et al}.} have tried
  to do. We admit that some selected special 
 properties like the long-wave susceptibility
 at $q=0$, if treated properly by a HTE, could be  
 approximated this way. 
 But quantities (e.g.\  $\chi({\bf q},\omega)$) with the ${\bf q}$-wave vector in 
 the vicinity of the 1D 'propagation vector' 
 defined by the maximum of the 1D
 magnetic 
 structure factor $S({\bf q},0)$) certainly do not belong to them. 
 This is clearly evidenced by 
 the magnon minimum (soft spin excitation 
 above about 2.5~meV, only, from the ground state) as
 seen in the INS \cite{Lorenz09} reflecting the vicinity  
 to a critical point. Thus, the low-energy
 dynamics of both models are  different
 since the microscopic mechanisms and the stability of the FM inchain
 ordering are completely distinct 
 for these two approaches.

  We hope that our detailed discussion of $\chi(T)$ at intermediate $T$ 
  is of interest to a broader readership.
  We have drawn attention to the available HTE computer codes. Thereby we stress that the usual 
  smooth behavior of $ \chi (T) $ doesn't allow one to deduce multiple exchange parameters, 
  especially if the maximum and the submaximum regions at low-$T$ are not 
  involved in the fit. Other thermodynamic 
  quantities such as the saturation field,
 magnetization, specific heat, and INS data must be included.
 What can and must be done when dealing with $\chi(T)$, 
 is to check the compatibility of already  existing derived or proposed
 parameter sets with experimental $\chi (T)$ data.  
 The model proposed by Shu {\it et al.} has not passed 
 our $\chi$-check.
 Their argument against our FM $\Theta_{\rm \tiny CW}\approx 50$~K \cite{Lorenz09} due to a seemingly
 ``incomplete selection of fitting parameters" is obsolete since a weak $J_a$ value
 has been included in our refined analysis resulting now in 
 $\Theta_{\rm \rm \tiny CW}\approx 55$~K.
 The stressed seemingly divergence of $\chi(T)$ at $T=\Theta_{{\rm \tiny CW}}\approx 50$~K
  of our frustrated set doesn't reflect 
 "an inconsistency" of our approach as claimed by Shu {\it et al.}. Rather, it
 provides just a proof and documents 
 only that the authors did  not consider
 the restricted
 regions of validity for {\it any} CW-law approaching even intermediate and of course lower $T$! 
In this context, we mention that, if the authors would have inserted
 their own  correct  sign values for $\Theta_{\rm \tiny CW}>0$ derived from its 
 definition by equation (4) (see table 1), they would
 be confronted themselves with the same pseudo-problem they have tried to accuse us!
 Their referring to the work \cite{Xiang07}\footnote{Here,ignoring any IC due to its seemingly
 weakness  the origin of the FM inchain ordering, instead of the expected spiral ordering for
 $\alpha >1/4 $  and the final $T_N$, has been attributed to an 'order from 
 disorder'-transition.} as a `first principle
density functional theory (DFT)' explanation for why a 
spiral ordering is missing, is somewhat misleading. There the 
 relative weak
 frustrating AFM IC's and the related and 
 observed \cite{Lorenz09} dynamical spiral fluctuations 
 were
ignored 
  but 
 Xiang {\it et al.} \cite{Xiang07}
 arrived
 nevertheless at a {\it non-negligible}
  finite $J_2$ value {\it above} $\alpha_c=1/4$, in qualitative accord with the INS data \cite{Lorenz09}. 
  To the best of our knowledge,
  there is no edge-sharing chain cuprate with a vanishing $J_2$ and a {\it finite} 
  frustrating AFM value is considered to be generic
  for the whole family. There is no reason to prescribe 
  a seemingly vanishing $J_2$ value to the presence
  of a significant level of O vacancies in high-quality samples. 
  The INS \cite{Lorenz09,Lorenz11} and  RIXS data \cite{Money16} 
  provide clear evidence for a skew AFM IC with no 
  need for a speculative and exotic `order from disorder' scenario.

 To summarize, 
 the frustrated model for Li$_2$CuO$_2$, 
 refined 
 by
 the INS data
 \cite{Lorenz09,Lorenz11,Boehm98} (s.\ the last row in table 1), our  DFT and
 five-band $pd$-Hubbard model calculations \cite{Lorenz09}, are consistent with the
 experimental data, including the
 $\chi (T)$ \cite{Lorenz09}, the field
 dependent magnetization, the saturation field 
 value
 \cite{Nishimoto11}, and the $T$-dependent RIXS spectra
 \cite{Money13}. This is in contrast to  the SLT-model,
 which does not account for these data.
Due to lacking space a critical discussion of the O vacancy scenario
 and the related Cu 4$p$ hole bonding scenario by Shu {\it et al.} 
instead of the correlated Cu 3$d$ covalent electron picture
 must be considered elsewhere.
 We mention here only
 that the main 
 reason for disorder in Ca$_2$Y$_2$Cu$_5$O$_{10}$
 is likely 
 {\it  not} 
 the 
 presence of 
 O vacancies, as stated by Shu {\it et al.}, but the intrinsic 
 misfit
 with the 
Ca$^{2+}$Y$^{3+}$-chains that surround and distort the CuO$_2$ chains 
\cite{Gotoh06,Thar06,Matsuda17}. 
Based on
this complex structure there are
 no 
 hints pointing to O vacancies.

\emph{Note on the Authors's Reply  and Corrigendum}\\
\indent {\it (for the convenience of the readers  of ArXiv).}
\begin{enumerate}
  \item In their ``Corrigendum'' \cite{Shu18cor} submitted later on
 but  published nevertheless shortly before our
Comment and the Authors's Reply \cite{Shu18},  Shu {\it et al.} have recognized 
that their equation (7) is wrong. In the proposed corrected equation, they 
use the notations of Ref.\ \cite{Takeda95} for the exchange couplings, i.e.\
the exchange interaction of a pair of spins is defined as $-2J^{\ast}$, in contrast to
our (and that of \cite{Shu17}) notations, where it reads $+J$. As we have already 
mentioned
above, a positive sign of $J$ corresponds to an AFM interaction. 
  \item Unfortunately, in the Reply \cite{Shu18},  Shu {\it et al.} repeat
  their erroneous statement that a positive value of $\Theta_{CW}$ would be allowed only
  for a system with an FM ground state (see the end of p.\ 5). Just for illustration, 
  let us consider the case of a
  simple cubic magnetic lattice with a large NN FM interaction along 
  the $x$-direction and a tiny NN AFM coupling along the $y$- and $z$-directions. It is 
  clear that the system will have a $\Theta _{CW}>0$ and an AFM ground state.  
    \item On pages 5,6 of Ref.\  \cite{Shu18},  Shu {\it et al.}
    erroneously claim that
    HTE (they call it ``HTSE'') would be an approximation to the Curie-Weiss law
    (see the paragraph below Fig.\ 3 of Ref.\  \cite{Shu18}). Moreover, they 
    state:
    ``Finally, the polynomial HTSE fitting can always achieve high 
    goodness-of-fit for nearly any curve without abrupt turns via higher 
    order parameter adjustment within a limited temperature range''.
    We may repeat here only that the high-temperature series
    expansion (HTE or HTSE) is an \emph{exact} expansion and that the Curie-Weiss law is
given by the first two terms of it (see equation (27) of section IV.B in \cite{Schmidt01}). 
   It is \emph{not a result} of the mean-field approximation. The terms 
   of any order of a HTE are \emph{calculated} from the Heisenberg model 
   and are unambiguously expressed via the exchange parameters {\it without}
    any adjustment. 
   \item In section 4.5 of  \cite{Shu18},  Shu {\it et al.} erroneously 
   claim that $\sum_{i<j}J_{ij}S_iS_j \neq \frac 12\sum_{i,j}J_{ij}S_iS_j$ and
   doubt that we properly interpret our exchange values. We may mention that
   in the notations of Eq.\ (\ref{eq:H}), the total width of the dispersion of
   spin excitations is $W\approx 2|J_1|$ \cite{Lorenz09}. The recently found 
   full magnetic dispersion for Ca$_2$Y$_2$Cu$_5$O$_{10}$ \cite{Matsuda14,Matsuda17}
   amounts $W\approx 55$~meV. So, the $J_1/k_B > 200$~K value 
   (that give $\Theta_{CW}\approx +80$~K) which is proved experimentally.
\end{enumerate}

\ack 
ROK thanks the  Heidelberg University for hospitality.
Also the project   III-8-16 of the NASc of the Ukraine and the NATO  project SfP 984735
are acknowledged. SN thanks the SFB 1143  of the DFG for financial support.
MM acknowledges the support by the Scientific User Facilities Division, 
Office of Basic Energy Sciences, US DOE.
We appreciate discussions with A.\ Tsirlin,  
B.\ B\"uchner, V.\ Grinenko, and
K. W\"orner. The publication of this article was funded by the Open 
Access Fund of the Leibniz Association.

\section*{ORCIDiDs}
R~Klingeler (id) https://orcid.org/0000-0002-8816-9614\\
R~O~Kuzian (id) https://orcid.org/0000-0002-6672-7224\\
C~Monney (id) https://orcid.org/0000-0003-3496-1435\\
\section*{References}

\bibliographystyle{iopart-num}
\bibliography{shu15}

\end{document}